# Societal Implicit Memory and his Speed on Tracking Extrema in Dynamic Environments using Self-Regulatory Swarms

**Vitorino Ramos[1]**, **Carlos Fernandes[2,3]**, **Agostinho C. Rosa[2]**

[1] CVRM-IST, Technical Univ. of Lisbon (IST),
Av. Rovisco Pais, 1, 1049-001, Lisbon, PORTUGAL
vitorino.ramos@alfa.ist.utl.pt
[2] LaSEEB-ISR-IST, Technical Univ. of Lisbon (IST),
Av. Rovisco Pais, 1, TN 6.21, 1049-001, Lisbon, PORTUGAL
{cfernandes,acrosa}@laseeb.org
[3] EST-IPS, Setúbal Polytechnic Institute (IPS),
R. Vale de Chaves - Estefanilha, 2810, Setúbal, PORTUGAL

*Abstract*— In order to overcome difficult dynamic optimization and environment extrema tracking problems, we propose a *Self-Regulated Swarm* (*SRS*) algorithm which hybridizes the advantageous characteristics of *Swarm Intelligence* as the emergence of a societal environmental memory or cognitive map via collective pheromone laying in the landscape (properly balancing the exploration/exploitation nature of the search strategy), with a simple *Evolutionary mechanism* that trough a direct reproduction procedure linked to local environmental features is able to self-regulate the above exploratory swarm population, speeding it up globally. In order to test his adaptive response and robustness, we have recurred to different dynamic multimodal complex functions as well as to *Dynamic Optimization Control* (*DOC*) problems. Measures were made for different dynamic settings and parameters such as, environmental upgrade frequencies, landscape changing speed severity, type of dynamic (linear or circular), and to dramatic changes on the algorithmic search purpose over each test environment (e.g. shifting the extrema). Finally, comparisons were made with traditional *Genetic Algorithms* (*GA*) as well as with more recently proposed *Co-Evolutionary* approaches. *SRS*, were able to demonstrate quick adaptive responses, while outperforming the results obtained by the other approaches. Additionally, some successful behaviors were found: *SRS* was able not only to achieve quick adaptive responses, as to maintaining a number of different solutions, while adapting to new unforeseen extrema; the possibility to spontaneously create and maintain different subpopulations on different peaks, emerging different exploratory corridors with intelligent path planning capabilities; the ability to request for new agents over dramatic changing periods, and economizing those foraging resources over periods of stabilization. Finally, results prove that the present *SRS* collective swarm of bio-inspired agents is able to track about 65% of moving peaks traveling up to ten times faster than the velocity of a single ant composing that precise swarm tracking system. This emerged behavior is probably one of the most interesting ones achieved by the present work.

## I. INTRODUCTION

MOST research in evolutionary (*EC*) and swarm intelligence (SI) computation focuses on optimization of static, non-changing problems. Many real-world optimization problems, however, are dynamic, and optimization is needed that are capable of continuously adapting the solution to a changing environment. In fact, many real-world problems are actually dynamic: new jobs have to be added to the schedule, machines may break down or wear down slowly, raw material is of changing quality, etc [8]. If the optimization problem is dynamic, the goal is no longer to find the extrema, but to track their progression through the space as closely as possible. One method for achieving this is to evolve a function off-line that models the dynamics of the environment directly [58] while another is to use the dynamics of the evolutionary -or Swarm Intelligent- process itself on-line to track the progression of the extrema. As described by *Angeline* [1] evolving a function that describes extrema dynamics is preferable when the dynamics can be modeled accurately off-line. When a single function cannot describe the dynamic accurately enough, an on-line approach using the implicit dynamics of the self-adaptive method is preferable. However, while evolutionary computation methods that evolve mutation variances perform well in static environments, it is not clear that these methods are beneficial when the gradient at each point is constantly in flux as in most dynamic environments [1], apart from significant, well-know and promising attempts. For these reasons, it seems appropriate to keep the suggestion of on-line methods, simultaneously attempting it with different computational paradigms such as those derived from Swarm Intelligence which allows for distributed real-time self-organization of solutions while maintaining strong adaptive capabilities. Following this new research path, we propose a *Self-Regulated Swarm* (*SRS*) algorithm which hybridizes the advantageous characteristics of Swarm Intelligence as the emergence of a societal environmental memory or cognitive map [50,49] via collective pheromone laying in the landscape (balancing the exploration/exploitation nature of the search strategy), with a simple evolutionary mechanism that trough a direct reproduction procedure linked to local environment features is able to self-regulate the above exploratory swarm population, speeding it up globally [17]. Our present proposal is fully discussed in section II.

### A. Approaches

Dynamic Optimization (*DO*) problems, on a more abstract level can be characterized by several situations; this might mean that the optimization function, the problem instance, or some restrictions may change, and thus the optimum to that problem might change as well. If any of these events are to be taken into account in the optimization process, we call the problem dynamic or changing (both terms are used synonymously and often in the literature, also the term non-stationary is used).

Since many approaches are possible, *Branke* and *Schmek* [8] suggested in 2002, a classification of *DO* problems, surveying it, while classifying a number of the most widespread techniques that have been published in the literature so far to make evolutionary algorithms (*EAs*) suitable for changing optimization problems. As pointed by them, one standard approach to deal with these dynamics is to regard each change as the arrival of a new optimization problem that has to be solved from scratch [48].

However, this simple approach is often impractical: solving a problem from scratch without reusing information from the past might be too time consuming, a change might not be identifiable directly, or the solution to the new problem should not differ too much from the solution of the old problem. Thus, as in the on-line tracking process suggested by *Angeline* [1], *Branke* [8,9,10] recommended that would be nice to have an optimization algorithm that is capable of continuously adapting the solution to a changing environment, reusing the information gained in the past. Since in nature adaptation is a continuous and continuing process, and *EAs* have much in common with natural evolution, they seem to be a suitable candidate. However, these type of evolutionary approaches typically converge to an optimum and thereby lose their diversity necessary for efficiently exploring the search space and consequently also their ability to adapt to a change in the environment when such a change occurs [8,10]. The problem here can be stated as seeking an appropriate and difficult balance between two contradictory characters of the search procedure; those between the *exploring* (ideal for gathering new solutions) and *exploiting* (making the best use of past solutions) nature of the algorithm.

Over the past few years, a number of authors have addressed the problem of convergence and subsequent loss of adaptability in many different ways. According to *Branke* and *Schmeck* [8], most of these approaches could be grouped into one of the following three categories established by them:

***I*** - *React on Changes*: The EA is run in standard fashion, but as soon as a change in the environment has been detected, explicit actions are taken to increase diversity and thus facilitating the shift to the new optimum.

***II*** - *Maintaining Diversity throughout the run*: Convergence is avoided all the time and it is hoped that a spread-out population can adapt to changes more easily.

***III*** – *Memory-based Approaches*: The EA is supplied with a memory to recall useful information from past generations, which seems especially useful when the optimum repeatedly returns to previous locations.

Techniques such as *Hypermutation* (*Cobb*, [13]) or *Variable Local Search* (VLS) (*Vavak*, [66,67]) pursue category ***I***, keeping the whole population after a change but increasing population diversity by drastically increasing the mutation rate for some number of generations. VLS on the other hand, increases mutation gradually after a change has been detected [66], where the mutation range itself can be adapted [67]. Another variant with significant improvements in convergence speed and solution quality have been found if the altered (old) individuals are reused (see *Bierwirth* [5,6], *Lin* [39], *Mattfeld* [41], or *Reeves* et al. [55]). *Artificial Neoteny* techniques proposed for *EAs* by *Ramos* [52] can also be considered among this approach. Neoteny, also know as *Paedomorphosis*, can be defined in biological terms as the retention by an organism of juvenile or even larval traits into later life. In some species, all morphological development is retarded; the organism is juvenilized but sexually mature. Such shifts of reproductive capability would appear to have adaptive significance to organisms that exhibit it, and the technique is used within *EAs* by simple re-injecting old artificial *DNA* which was randomly captured on the first generations. Finally, the ***A*** approach ends up with a number of authors suggesting to leave the response to a change in the environment up to the strategy itself, by relying on his *self-adaptiveness*. This is the case in works by *Angeline* [1], *Bäck* [4], *Grefenstette* [20] and *Stephens* [60].

Maintaining diversity is another possible approach (***II***). *Grefenstette* [21] introduced the method *Random Immigrants* where in every generation, the population is partly replaced by randomly generated individuals, while *Andersen* [2] uses genotypic and phenotypic *sharing* to increase the Genetic Algorithm ability to track optima in slowly changing environments. A very interesting sharing scheme was introduced by *Liles* and *DeJong* [38], based on *Tag bits*. There, tag bits are appended to each genotype, and only individuals with equal tag bits are allowed to mate, which somehow can be regarded as introducing subpopulations of varying size [59]. The neighborhood used for sharing is then the number of individuals with the same tag bit, i.e. individuals with rare tag bits are favored. The approach is able to maintain different subpopulations on different peaks, however for a simple environment with two peaks of changing weights. Other techniques include a crowding-like replacement scheme entitled "*Worst among Most Similar*" by *Cedeno* [11], by taking in account each individual's age, modifying the fitness function $f_{mod} = g (f_{old}, age)$ as proposed by *Ghosh* [18] or finally by using the *Thermodynamical Genetic Algorithm* (TDGA) proposed by *Mori* [45] addressing the diversity in the population explicitly controlling a measure entitled "free energy" $F$. For a minimization problem this term is calculated as $F = <E>-TH$, where $<E>$ stands for the average population fitness and $H$ is a measure for the diversity in the population. The temperature $T$ is a parameter of the algorithm and reflects the emphasis of diversity (adjusting $T$ is discussed in [46], a problem similar in many respects in what we found to control scheduling temperatures in *Simulated Annealing*, e.g. [52]). Finally, a very interesting and recent approach was implemented by *Huang* and *Rocha* [30,56] outperforming traditional Genetic Algorithms via obtaining greater phenotypic plasticity. They proposed a co-evolutionary agent based model of genotype editing (*ABMGE*), constructed using several genetic editing characteristics that are gleaned from the RNA editing system as observed in several organisms. The incorporation of editing mechanisms provides a means for artificial agents with genetic descriptions to gain greater phenotypic plasticity. By allowing the family of editors and the genotypes of agents to co-evolve using the re-generation of editors as a control switch for environmental changes, the artificial agents in *ABMGE* can discover proper editors to facilitate the tracking of the extrema in dynamic environments. Other approaches on this line are fully discussed in [8,9,10] by *Branke* and *Schmeck*.

Another kind of approach is to supply the algorithm with some sort of memory (approach ***III***) storing good partial solutions in order to reuse them later. This can be advantageous in cases where the environment is changing periodically, and repeated situations occur. However they also could be counterproductive if the environment changes dramatically with open-ended novelty. Memory may be provided in two general ways: *implicitly* by using redundant representations, or *explicitly* by introducing an extra memory and formulating strategies to deposit and retrieve solutions later. Generally, the most prominent approach to implicit memory and redundant representation is multiploidy [19] (e.g., *Ryan* [57], *Lewis* [37] and *Dasgupta* [16]). On the other hand, while redundant representations allow the *EA* to implicitly store some useful information during the run, it is not clear that the algorithm actually uses this memory in an efficient way. As an alternative, the following approaches use an explicit memory in which specific information is stored and reintroduced into the population at later generations. The memory can be used to transfer individuals from one *EA* run to seed the initial population on another *EA* after a single change in the environment, as suggested by *Louis* and *Xu* [40], or trough the use of a knowledge base to memorize successful individuals in a permanent memory, assuming that the system can measure the environmental conditions (*Ramsey* and *Grefenstette* [54]). Other examples include a procedure by *Trojanowski* [64] where each individual is extended with additional memory for a number of its ancestors, or via a variation of evolutionary elitism within *Thermodynamical Genetic Algorithms* [44,45,46] – i.e., in every generation the best individual is stored in the memory, and another individual is deleted from the memory depending on its age and contribution to this memory population's diversity (measured as variance over bit positions). Finally, *Branke* [9] compared a number of replacement strategies for inserting new individuals into a memory stressing the importance of diversity for memory-based approaches. As an example, *Branke* found out that a simple replacement of the most similar individual performed almost equivalently to a strategy that replaces the worse of the two individuals in the memory closest to each other.

Although the memory/search population approach was quite successful, it soon became obvious that a strategy based on memorization would be too restricted to adapt successfully to a wide range of dynamic environments. As an alternative, *Branke* and *Schmek* [8] developed an approach with multiple populations acting as *Self-Organizing Scouts* (*SOS*), watching over the changing landscape. The basic idea of *SOS* is that once a peak has been found (i.e. the population converged to a high-performance region), the population should split, while the "child population" should "watch" over that peak, the remainder should spread out searching for new peaks. Since our current approach is also based in Self-Organization [50,53], trough the use of stigmergic mechanisms as well as *Swarm Intelligence* [7], there are in fact some similarities as well differences between the *Self-Regulated Swarms* SRS approach and SOS that will be interesting to discuss later (see section V-B).

Finally some recent proposals have been made using a Swarm Intelligent (SI) [7, 33] approach to attempt to solve these dynamic problems. Generally, *Swarm Intelligence* can be regarded as the property of a system whereby the collective behaviors of (unsophisticated) entities interacting locally with their environment cause coherent functional global patterns to emerge. SI provides a basis with which it is possible to explore collective (or distributed) problem solving without centralized control or the provision of a global model (Stan Franklin, *Coordination without Communication*, talk at Memphis Univ., USA, 1996) [49]. These entities can be either regarded as bio-inspired ant-like agents in which self-organization occurs trough trail formation via pheromone deposition and evaporation [7,50,17,49,23,53,51] giving rise to the well know *Ant Colony Systems* (*ACS*) and *Ant Colony Optimization* (*ACO*) algorithms by *Dorigo* et al. [7], or as physical particles embodied with direction, velocity and intrinsic memory for best global and local position [33,2,3,15,32] know as *Particle Swarm Optimization* (PSO) algorithms developed by *Kennedy*, *Eberhart* et al. [33].

Apart from the SI research branch taken (*ACS* or *PSO*), only very recently there are being made efforts for dynamic optimization problems using Swarm Intelligence. Via *ACS*, *Guntsch* and *Middendorf* [23], applied population based *ACO* algorithms for tracking extrema in dynamic environments, while others like *Ramos*, *Fernandes* and *Rosa* [50,17,49] developed distributed pheromone laying over the dynamic environment itself, in order to track different peaks. They show that the self-organized algorithm is able to cope and quickly adapt to unforeseen situations even when over the same cooperative foraging period, the community is requested to deal with two different and contradictory purposes (different extrema), comparing their results with those from *Passino* et al. [47] which developed a *Bacterial Foraging Optimization Algorithm* (*BFOA*) for distributed optimization and control. Finally also via *ACS*, *Guntsch*, *Middendorf* and *Schmeck* [22] developed an *ACO* introducing local variance where needed, and a heuristic repair of solutions. They applied successfully the novel algorithm for the combinatorial dynamic TSP (*Travelling Salesman Problem*).

Meanwhile via the *PSO* [33] approach, *Carlisle* and *Dozier* [2,3], adapted it for dynamic environments. The process consists of causing each particle in the swarm to reset its record of its best position as the environment changes, to avoid making direction and velocity decisions on the basis of outdated information. Thus, the authors explicitly discard memory from the standard *PSO*, and call it *Adaptive PSO*. On the other hand, *Parrott* and *Li* [15] developed a *PSO* model for tracking a three-peak multimodal environment. To achieve this, a form of speciation allowing development of parallel subpopulations is used. The model employs a mechanism to encourage simultaneous tracking of multiple peaks by preventing overcrowding at peaks. Others like *Janson* and *Middendorf* [32], followed an hierarchical strategy adjusting the *PSO* accordingly and developing the first hierarchical *PSO* for dynamic problems.

Other authors like *Altshuler* and *Wagner* [69], presented a multi-agent system for a dynamic cleaning problem on the area of *Swarm Cooperative Robotics*. In order to find the minimal cleaning time possible for a given "contaminated" shape, they have designed the system in such a way that all the cleaning agents are controlled by a central unit (referred to as the *queen*). Upon initialization, the queen is given the complete information regarding the contaminated shape to be cleaned. While the agents are traveling along the grid, the queen is immediately aware of any new information discovered by the agents. The queens orders as to the next desired movements of the agents are also immediately transferred to the agents, which carry them our automatically.

### B. Dynamism Characterization and Benchmarks

*Branke* and *Schmeck* [8,10] suggested a number of criteria along which dynamic environments could be categorized as well as tested. Since many other authors on dynamic optimization and control are following them, we will adopt them as well, having in mind different comparison purposes. Four main aspects should be considered. The first two are usually adopted:

*I – Frequency of change*: This criterion establishes how often the environment changes (starting from very rare changes up to continuous change). A parameter *uf* (*update frequency*) is introduced. If for instance *uf = C*, then at every *C* iterations (or generations, in case of *EA's*) the landscape will suffer changes.

*II – Severity of change*: This criterion establishes how strongly the system is changing. From slight changes to completely new situations. The severity parameter *S* will be defined later.

*III – Predictability of change*: The present aspect gives a notion if there is a pattern or trend in the changes. That is, depending on the problem at hands, it is somehow possible to predict direction, time or, the severity of the next change given the changes encountered so far?

*IV – Cycle length*: This criterion measures how often the optimum returns to previous locations or at least gets close to them.

Since it seems difficult to measure some of these characteristics, comparisons between different optimization problems may still be out of reach, but at least the above characteristics can be varied on a single problem such that their qualitative influence on a specific approach can be examined [8]. In our different experiments (section III) we will use criterions of type IV (section III.A), type I (section III.B) and of type II (section III.C), or with combinations with the formers. Moreover, in order to compare the relative abilities of different algorithmic approaches for on-line tracking of dynamic extrema, *Angeline* [1] suggested a number of dynamical environments created from a single multi-dimensional function, referred to as *the base function* below. Dynamic environments were then obtained by translating the base function along a number of distinct temporal trajectories (fig. 1). This setup allows complete control of the dynamics and generates simple yet non-trivial dynamic functions with know properties. Tests under this specific benchmark were performed in section III.A. *Angeline* [1] has made use of a simple convex parabolic function in three dimensions, while we tested our algorithm over two complex multimodal functions, well know in evolutionary computation performance evaluation: the *Ackley* (fig. 3) and *Schaffer* F7 functions. The *Ackley* function was used as our base function for section III.A tests.

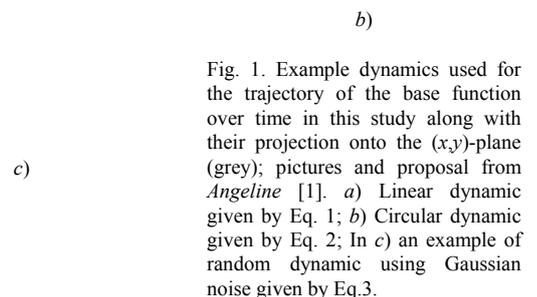

Fig. 1. Example dynamics used for the trajectory of the base function over time in this study along with their projection onto the (*x,y*)-plane (grey); pictures and proposal from *Angeline* [1]. *a*) Linear dynamic given by Eq. 1; *b*) Circular dynamic given by Eq. 2; In *c*) an example of random dynamic using Gaussian noise given by Eq.3.

As described by *Angeline*, the temporal dynamics applied to the base function involve three distinct parameters: *dynamic type*, *step size* and *update frequency*. Dynamic type identifies one of three distinct parameterized methods for translating the base function. Severity is a parameter used as input to each of the dynamics to determine the amount the base function is displaced from its current position. Finally, update frequency determines the number of iterations (or generations) between each movement of the base function. *Angeline* [1] tested three different dynamic types in his experiments: linear, circular and random. Given a severity of *S*, the linear dynamic updates the current offset for dimension *k*, denoted as $\Delta'_k$, as follows (Eq.1):

$$\Delta'_k = \Delta_k + S \qquad (1)$$

which simply adds a constant displacement to the offset in each dimension equivalent to the severity parameter's setting. On the other hand, the circular dynamic is computed as follows (Eq.2):

$$\Delta'_k = \Delta_k + S \cdot \sin\left(\frac{2\pi t}{C}\right), k = 1,3$$
$$\Delta'_k = \Delta_k + S \cdot \cos\left(\frac{2\pi t}{C}\right), k = 2 \qquad (2)$$

where *t* is the number of applications of the dynamic thus far. Equation 2 describes a trajectory that translates the base function in a circular path through all three dimensions. The equation is designed to cycle the values of the offset every *C* applications with the severity parameter determining the radius of the circular path. Finally, for the random dynamic, random noise is added to the offset as follows (Eq. 3):

$$\Delta'_k = \Delta_k + S \cdot N(0,1) \qquad (3)$$

where *N*(0,1) is a Gaussian random variable with mean 0 and variance 1. Here, the severity parameter determines the variance of the noise added to the base function offset. Figure 1 (from [1]) shows an example of a random trajectory generated by this dynamic. In our tests (section III) we have made use of linear and circular dynamics (Eq.1,2) using the *Ackley* multimodal complex function (fig.3) as our base function.

### C. Our Proposal

Many structures built by social insects are the outcome of a process of self-organization [50,49,51], in which the repeated actions of the insects in the colony interact over time with the changing physical environment to produce a characteristic end state [25]. A major mediating factor is stigmergy [63], the elicitation of specific environment-changing behaviors by the sensory effects of local environment changes produced by previous and past behavior of the whole community. Stigmergy is a class of mechanisms that mediate animal-animal interactions through artifacts or via indirect communication, providing a kind of environmental synergy, information gathered from work in progress, a distributed incremental learning and memory among the society. In fact, the work surface is not only where the constituent units meet each other and interact, as it is precisely where a dynamical cognitive map could be formed, allowing for the embodiment of emergent adaptive memory, cooperative learning and perception [50,49]. Constituent units not only learn from the environment as they can change it over time. Its introduction in 1959 by Pierre-Paul Grassé[1] made it possible to explain what had been until then considered paradoxical observations: In an insect society individuals work as if they were alone while their collective activities appear to be coordinated. The stimulation of the workers by the very performances they have achieved is a significant one inducing accurate and adaptable response (check applications in [53,51]). Keeping in mind these characteristics we will present a stigmergic self-regulated model to tackle the collective adaptation of a social swarm for dynamic tracking, based on real ant colony behaviors.

### II. SELF-REGULATED SWARMS; THE *SRS* ALGORITHM

As mentioned above, the distribution of the pheromone represents the memory of the recent history of the swarm (his social cognitive map), and in a sense it contains information which the individual ants are unable to hold or transmit [50]. There is no direct communication between the organisms but a type of indirect communication through the *pheromonal* field.

In fact, ants are not allowed to have any local memory and the individual's spatial knowledge is restricted to local information about the whole colony pheromone density. In order to design this behaviour, one simple model was adopted [12], and extended due to specific constraints of the present proposal, in order to deal with 3D dynamic environments. As described by *Chialvo* and *Millonas*, the state of an individual ant can be expressed by its position *r*, and orientation *θ*. Since the response at a given time is assumed to be independent of the previous history of the individual, it is sufficient to specify a transition probability from one place and orientation (*r*,*θ*) to the next ($r^*,\theta^*$) an instant later. In a previous works by *Millonas* [42,43], transition rules were derived and generalized from noisy response functions, which in turn were found to reproduce a number of experimental results with real ants. The response function can effectively be translated into a two-parameter transition rule between the cells by use of a pheromone weighting function (Eq.4):

$$W(\sigma) = \left(1 + \frac{\sigma}{1+\gamma\sigma}\right)^{\beta} \qquad (4)$$

This equation measures the relative probabilities of moving to a cite *r* (in our context, to a cell in the grid *habitat*) with pheromone density *σ*(*r*). The parameter *β* is associated with the osmotropotaxic sensitivity, recognised by *Wilson* [68] as one of two fundamental different types of ant's sense-data

---

[1] Grassé, P.P.: La reconstruction du nid et les coordinations inter-individuelles chez *Bellicositermes natalensis* et *Cubitermes sp*. La théorie de la stigmergie : Essai d'interpretation des termites constructeurs. *Insect Sociaux* (1959), 6, 41-83. The phrasing of his introduction to the term stigmergy is worth noting (translated to English in [OK30]): *The coordination of tasks and the regulation of constructions do not depend directly on the workers, but on the constructions themselves. **The worker does not direct his work, but is guided by it**. It is to this special form of stimulation that we give the name Stigmergy (**stigma** - wound from a pointed object, and **ergon** - work, product of labor = stimulating product of labor)*.

TABLE I
HIGH-LEVEL DESCRIPTION OF THE **SELF-REGULATED SWARM** (*SRS*) ALGORITHM PROPOSED

```
/* Initialization */
For all ants do
   Place ant at randomly selected site r
   e[ant]=1.0
End For

/* Main loop */
For t = 1 to t_max do
   For all ants do
      /* According to Eqs. 4 and 5 (section II) */
      Compute W(σ) and P_ik
      Move to a selected neighboring site not occupied by other ant
      /* According to Eq. 6 (section II) */
      Increase pheromone P_r at site r: P_r= P_r+[η+p(Δ[r]/Δmax)]
      /* Reproduction procedure (section II - A) */
      Compute n, the number of occupied surrounding cells
      If ant meets ant (i.e., n ≥ 1) then
         Determine P**(n)
         Compute reproduction probability P* = P**(n) [Δ(r)/Δmax]
         If real random [0, 1] < P* then
            Create one ant with e[ant]=1.0 and place it randomly on one of
                the free cells surrounding the main parent at site r
         End If
      End If
   End For
   Evaporate pheromone by K, at all grid sites
   For all ants do
      Decrease ant energy: e[ant]=e[ant]- Δe
      If e[ant] ≤ 0.0 then
         Kill that ant
      End If
   End For
   Print location of all agents
   Print pheromone distribution at all sites
End For

/* Values of parameters used in experiments */
k = 1.3, η = 0.07, β=3.5, γ=0.2, Δe=0.1,
p = 1.9, t_max = 100 or 400 time steps.
/* Constant values */
P**(0) = P**(8) =0, P**(4) = 1, P**(5) = P**(3) =0.75,
P**(6) = P**(2) =0.5, P**(7) = P**(1) = 0.25
/* Useful references */
Check [50], [17], [49], [53] and [12].
```

processing. *Osmotropotaxis*, is related to a kind of instantaneous pheromonal gradient following, while the other, *klinotaxis*, to a sequential method (though only the former will be considered in the present work as in [12]). Also it can be seen as a physiological inverse-noise parameter or gain.

In practical terms, this parameter controls the degree of randomness with which each ant follows the gradient of pheromone. On the other hand, $1/\gamma$ is the sensory capacity, which describes the fact that each ant's ability to sense pheromone decreases somewhat at high concentrations.

In addition to the former equation, there is a weighting factor $w(\Delta\theta)$, where $\Delta\theta$ is the change in direction at each time step, i.e. measures the magnitude of the difference in orientation. As an additional condition, each individual leaves a constant amount $\eta$ of pheromone at the cell in which it is located at every time step $t$.

This pheromone decays at each time step at a rate $k$. Then, the normalised transition probabilities on the lattice to go from cell $k$ to cell $i$ are given by $P_{ik}$ (Eq. 5, [12]), where the notation $j/k$ indicates the sum over all the surrounding cells $j$ which are in the local neighbourhood of $k$. $\Delta_i$ measures the magnitude of the difference in orientation for the previous direction at time $t$-1. Since we use a neighbourhood composed of the cell and its eight neighbours, $\Delta_i$ can take the discrete values 0 through 4, and it is sufficient to assign a value $w_i$ for each of these changes of direction. *Chialvo* et al, used the weights of $w_0$ =1 (same direction), $w_1$ =1/2, $w_2$ =1/4, $w_3$ =1/12 and $w_4$ =1/20 (U-turn). In addition coherent results were found for $\eta$=0.07 (pheromone deposition rate), $k$=0.015 (pheromone evaporation rate), $\beta$=3.5 (osmotropotaxic sensitivity) and $\gamma$ =0.2 (inverse of sensory capacity), where the emergence of well defined networks of trails were possible. Except when indicated, these values will remain in the following test framework. As an additional condition, each individual leaves a constant amount

$$P_{ik} = \frac{W(\sigma_i)w(\Delta_i)}{\sum_{j/k} W(\sigma_j)w(\Delta_j)} \quad (5)$$

$$T = \eta + p\frac{\Delta[i]}{\Delta_{max}} \quad (6)$$

$\eta$ of pheromone at the cell in which it is located at every time step $t$. Simultaneously, the pheromone evaporates at rate $k$, i.e., the pheromonal field will contain information about past movements of the organisms, but not arbitrarily in the past, since the field *forgets* its distant history due to evaporation in a time $\tau \cong 1/k$. As in past works, toroidal boundary conditions are imposed on the lattice to remove, as far as possible any boundary effects (e.g. one ant going out of the grid at the south-west corner, will probably come in at the north-east corner).

In order to achieve emergent and *autocatalytic* mass behaviours around specific extrema locations (e.g., peaks or valleys) on the *habitat*, instead of a constant pheromone deposition rate $\eta$ used in [12], a term not constant is included. This upgrade can significantly change the expected ant colony cognitive map (pheromonal field). The strategy follows an idea implemented earlier by *Ramos* [53], while extending the *Chialvo* model into digital image habitats, aiming to achieve a collective perception of those images by the end product of swarm interactions. The main differences to the *Chialvo* work is that ants, now move on a 3D discrete grid, representing the functions which we aim to study (figs. 3,4, section III) instead of a 2D *habitat*, and the pheromone update takes in account not only the local pheromone distribution as well as some features of the cells around one ant. In here, this additional term should naturally be related with specific characteristics of cells around one ant, like their altitude ($z$ value or function value at coordinates $x,y$), having in mind our present aim. So, our pheromone deposition rate $T$, for a specific ant, at one specific cell $i$ (at time $t$), should change to a dynamic value ($p$ is a constant = 1.93) expressed by equation 6. In this equation, $\Delta_{max} = | z_{max} - z_{min} |$, being $z_{max}$ the maximum altitude found by the colony so far on the function *habitat*, and $z_{min}$ the lowest altitude. The other term $\Delta[i]$ is equivalent to (if our aim is to minimize any given landscape): $\Delta[i] = |z_i - z_{max}|$, being $z_i$ the current altitude of one ant at cell *i*. If on the contrary, our aim is to maximize any given dynamic landscape, then we should instead use $\Delta[i] = |z_i - z_{min}|$. Finally, notice that if our landscape is completely flat, results expected by this extended model will be equal to those found by *Chialvo* and *Millonas* in [7], since $\Delta[i]/\Delta_{max}$ equals to zero. In this case, this is equivalent to say that only the swarm pheromonal field is affecting each ant choices, and not the *environment* - i.e. the expected network of trails depends largely on the initial random position of the colony, and in trail clusters formed in the initial configurations of pheromone. On the other hand, if this environmental term is added a stable and emergent configuration will appear which is largely independent on the initial conditions of the colony and becomes more and more dependent on the nature of the current studied dynamic *landscape* itself.

As specified earlier, the environment plays an active role, in conjunction with continuous positive and negative feedbacks provided by the colony and their pheromone, in order to achieve a stable emergent pattern, societal memory and distributed learning by the community [50,53].

### A. Reproduction procedure

In addition to the above advantageous characteristics of *Swarm Intelligence* as the emergence of a societal environmental memory or cognitive map [50,49] via collective pheromone laying in the dynamic landscape, we hybridized it with a simple evolutionary mechanism that trough a direct reproduction procedure linked to some local environmental features is able to self-regulate the above exploratory swarm population, speeding it up globally. The full **SRS** strategy adapted for dynamic extrema tracking then finally consists of using *Swarms with Varying Population Size* (SVPS) proposed and analyzed earlier by *Fernandes*, *Ramos* and *Rosa* [17].

This characteristic is achieved by allowing ants to reproduce and die through their evolution in the landscapes. To be effective, the process of variation must incorporate some kind of environmental pressure towards successful behavior, that is, ants that reach peaks/valleys must have some kind of reward, by staying alive for more generations – generating more offspring - or by simply having a higher probability of generating offspring at each time step. In addition, the population density in the area surrounding the parents must be taken into account during a reproduction event. When one ant is created (during initialization or by the reproduction process) a fixed energy value is assigned to it ($e[ant] = 1$). Every time step, the ant's energy is decreased by a constant amount of $\Delta e$ (usually 0.1). The ant's probability of survival after a time step is proportional to its energy during that same iteration, which means that after ten generations (with $\Delta e = 0.1$), this and other ants will inevitably die ($e[ant] = 0$). Within these settings one ant that is, for instance, 7 iterations old, has a probability of 0.3 to survive through the current time step. Meanwhile, for the reproduction process, we assume the following heuristic: an ant (main parent) triggers a reproduction procedure if it finds at least another ant occupying one of its 8 surrounding cells (*Moore* neighborhood is adopted). The probability $P^*$ (Eq.7) of generating offspring – one child for each reproduction event – is computed in two steps (see table I).

$$P^* = P^{**}(n) \left[ \frac{\Delta(r)}{\Delta_{max}} \right] \qquad (7)$$

First, the surrounding area is inspected in order to see if it is too crowded. Being *n* the number of occupied cells around this ant, the probability to reproduce $P^{**}$ is set to the values shown in table II. Notice that: 1) an ant completely surrounded by other ants, or isolated ($n=8$, $n=0$) do not reproduce; 2) the maximum probability is achieved when the area that ant is half occupied ($n=4$). After the probability $P^{**}$ is set to one of the previous values, the final probability is computed according to Eq.7, with the help of $\Delta(r)$ and $\Delta_{max}$ (similarly as in Eq. 6). This operation guarantees that any ant reaching the higher/lower peaks/valleys has more chance to produce offspring (notice that one ant in the higher/lower site has $P^{**}=1$ if $n=4$ and will reproduce for certain). If this ant passes the reproduction test (table I), then a new agent is created occupying one of his vacant cells around the main parent. None infant ants are allowed to be allocated in places where other ants are. Finally notice that when $n=0$ or $n=8$ no reproduction takes place and that higher/lower (maximization/ minimization) ants have more chance to reproduce.

TABLE II
REPRODUCTION PROBABILITY $P^{**}$ ACCORDING TO *N MOORE* NEIGHBORS

| *n* Moore neighbors | Reproduction probability $P^{**}(n)$ |
|---|---|
| $n = 0$ or $n = 8$ | 0.00 |
| $n = 4$ | 1.00 |
| $n = 5$ or $n = 3$ | 0.75 |
| $n = 6$ or $n = 2$ | 0.50 |
| $n = 7$ or $n = 1$ | 0.25 |

### B. Shifting the Extrema - Past work results

One of the features of *Swarms with Varying Population Size SVPS* discussed in [17] was the ability to adapt to sudden changes in the roughness of the landscape. These changes were simulated by abruptly replacing one test function by another after the swarm reached the desired regions of the landscape. Another way of simulating changes in the environment consists on changing the task from minimization to maximization (or vice-versa). The swarm performance was convincing and reinforced the idea that the system is highly adaptable and flexible. Further tests using *SVPS* concluded that varying population size increases the capability of the swarm to react to changing landscapes [17]. Figure 2 shows *SVPS* trying to find the lower values of *Passino F1* [47] until *t*=250, and then searching for the higher values. Comparisons with fixed sized swarms and *Bacterial Foraging Optimization Algorithms* (*BFOA*, [47]) were made in [49,50].

*a*) 3D view      *b*) 2D view

| $t = 10$ | $t = 250$ | $t = 260$ | $t = 280$ |
|---|---|---|---|
| $t = 300$ | $t = 320$ | $t = 500$ | |

Fig. 2. SVPS evolving on a complex multimodal function seen in *a-b*) [17,47,49]: the self-organized swarm emerges a characteristic flocking migration behavior between one deep valley (south region) and one peak (north region), surpassing in intermediate steps (*Mickey Mouse* shape at $t = 300$) some local optima. Over each foraging step, the population self-regulates. From $t=0$ to $t=250$ the swarm is induced to search the lowest valleys of the landscape. After $t=250$ the task changes (target peak moves to the north of the territory) and the swarm must find the higher values of the function. Check for detailed results and extended analysis in [17].

### III. DYNAMIC ENVIRONMENT TESTBED AND RESULTS

In order to compare the relative behavior of our present *Self-Regulated Swarms* (*SRS*) approach for on-line tracking of dynamic extrema, we followed diverse kind of test beds, reflecting different dynamism characterization and benchmarks described earlier in section I.B, namely according to: *dynamic type* (section III.A), *severity and speed tests* (section III.B), as well as the application of the current self-organized algorithm to *dynamic optimal control* problems (section IV).

#### A. Dynamic Type Tests

Over this specific test bed, dynamic environments were obtained by translating the *Ackley* complex multimodal base function (fig.3) along a number of distinct linear and circular temporal trajectories (fig.1). This setup allows complete control of the dynamics and generates simple yet non-trivial dynamic functions with know properties. The *Ackley* function [14,28] (Eq.8) is considered as a minimization problem. Originally this problem was defined for two dimensions, but the problem has been generalized to *N* dimensions [62]. Formally, this problem can be described as finding a string $x_i = \{x_1, x_2, …, x_N\}$, under the domain $x_i \rightarrow (-32.768, 32.768)$, that minimizes equation 8. In order to define an instance of this function we need to provide the dimension of the problem (in here, $n=2$). The optimum solution of the problem is the vector $u = (0,...,0)$ with $F(u)=0$. We will apply it to several swarm dynamic tests extending past analysis [50,17].

$$F\left(\vec{x}\right) = -20.\exp\left(-0.2\sqrt{\frac{1}{n}\sum_{i=1}^{n}(x_i - a_i)^2}\right) - \exp\left(\frac{1}{n}\sum_{i=1}^{n}\cos(2\pi.(x_i - a_i))\right) + 20 + e \quad (8)$$

For our specific tests we have made use of the domain $x,y \rightarrow [-2.0,2.0],[-2.0,2.0]$, containing 16 valleys. In order to change it over time, we have defined a target path line over our domain, from northwest (-2,2) to southeast (2,-2). Figure 3, shows three typical minimal target points along this line, *A*, *B* and *C*: *A*(-1.5;1.0), *B*(0.0;0.0) and *C*(1.0;-1.5). To introduce the dynamics into the problem, the base function minimal target point starts to move from *B*, moving continuously to southeast (passing *C*). Arriving at the southeast corner, the target point then moves to the northwest corner (since our *habitat* is toroidal), then passing *A*, and *B* again, over and over again has times passes, for several specific test speeds. Note that in equation 8, $a_i$ defines the point coordinates where the *Ackley* function has his minimal target point, thus introducing a more severe change than translation itself, since around that minimal point all the function changes differently depending on where this minimal point precisely is. This collection of dynamics when used with the complex multimodal *Ackley* base function has a number of nice features for testing the performance of self-adaptive algorithmic optimization. The linear dynamic will always move away from the position that the population is converging towards with the speed parameter determining how quickly it moves away. The *SRS* algorithm was then tested for different test speeds: $v=0$ (static environment), $v=0.5$, $v=1$, $v=1.5$, $v=2$, $v=3$, $v=5$ and $v=10$. In order to have an idea of the severity of this parameter, for instance at $v=2$, the target valley is traveling to southeast 2 habitat cells at each $t$ time step starting in the $B(0.0;0.0)$ mid-point, while each *SRS* ant can only move one cell for each time step $t$. *SRS* parameter values were the usual ones as indicated in table 1, section II.

| $z_{min}$=0 at $(x,y)$=-1.5,1 (normal view) | $z_{min}$=0 at $(x,y)$=-1.5,1 (from above) |
|---|---|
| $z_{min}$=0 at $(x,y)$=0,0 (normal view) | $z_{min}$=0 at $(x,y)$=0,0 (from above) |
| $z_{min}$=0 at $(x,y)$= 1,-1.5 (normal view) | $z_{min}$=0 at $(x,y)$=1,-1.5 (from above) |

Fig. 3. The *Ackley* complex multimodal function seen from different perspectives and with respective global minimum $z_{min}$=0 at *A*(-1.5;1.0), *B*(0.0;0.0) and *C*(1.0;-1.5), over the domain $x,y \rightarrow$ [-2.0,2.0],[-2.0,2.0] containing 16 valleys.

Figure 4 shows *SRS* agents space-time distribution (a-c) and their pheromone distribution (b-d) for $v=0$ (a-b) (the target valley is constantly at $B=(0;0)$) and for $v=0.5$ (c-d) (the valley is traveling to southeast one cell at each two time steps starting at the *B* mid-point), for several time steps $t=1,5,10,15,20, 30,50,70$ and 100. In (b-d) black pixels show where each ant agent is placed in the environment at a precise time step, starting initially by a random distribution at $t=0$. On the other hand, pheromone distribution represents the societal memory and the recent history of the swarm (considered as his emergent social cognitive map [12,49,50]), and in a sense it contains distributed information embedded within the dynamic environment itself. This pheromone distribution is represented in grey levels, i.e., points in space with higher levels of pheromone are represented by darker pixels. The mapping from the amount of pheromone at each cell and its correspondent gray level is linear and coded in 8 bit images (i.e., 256 gray levels). Figures 5, 6 and 7 shows *SRS* results for speeds $v=1$, $v=1.5$, $v=2$, $v=3$, $v=5$ and $v=10$. These tests and respective figures show that for low values of dynamic environment speed ($v=0$, $v=0.5$, $v=1$ and $v=2$), the swarm can consistently track the target extrema while he moves through the entire space. This is perfectly visible for speeds $v=0$ and $v=0.5$ (fig.4) where the target remains moving to the southeast corner over his 100 time steps run. As the target moves the swarm as a whole keeps following and converging to it, simultaneously self-regulating his population, i.e., downsizing it. At the same time, the whole swarm emerges a strong pheromone concentration at the right location of the target, increasing this concentration as times passes. For $v=1$ and $v=1.5$ however, due to speed, the target passes once at the southeast corner and reappears at $t=50$, 40 respectively for $v=1$, 1.5, at the northwest corner. In here, the swarm as a whole evolves a different global behavior. For $v=1$ and $t=50$ (fig.5), the swarm quickly adopts a short path (moving northeast) instead of moving ahead to northwest (where the target is), taking profit of the toroidal habitat. The same is true for $v=1.5$ and for $t=40$, whereas in here the swarm splits into two different clouds or clusters, one with the above short path strategy, while the other moving directly to it, following backwards the linear path ($NW \rightarrow SE$) of the target dynamic.

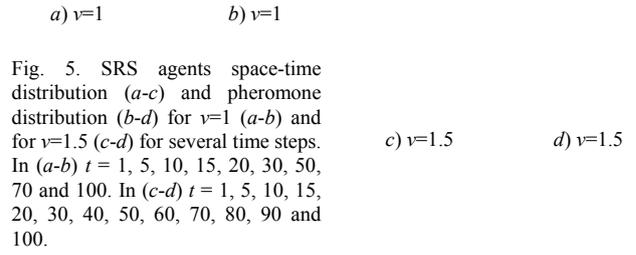

*a)* $v=1$     *b)* $v=1$     *c)* $v=1.5$     *d)* $v=1.5$

Fig. 5. SRS agents space-time distribution (*a-c*) and pheromone distribution (*b-d*) for $v=1$ (*a-b*) and for $v=1.5$ (*c-d*) for several time steps. In (*a-b*) $t$ = 1, 5, 10, 15, 20, 30, 50, 70 and 100. In (*c-d*) $t$ = 1, 5, 10, 15, 20, 30, 40, 50, 60, 70, 80, 90 and 100.

Every test started with a number of ant-like agents equal to ⅓ of the habitat size (100 x 100 cells) as in [12,53]. We see that generally for speed tests up to $v=5$, the population more or less

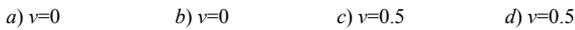

*a)* $v=0$     *b)* $v=0$     *c)* $v=0.5$     *d)* $v=0.5$

Fig. 4. SRS agents space-time distribution (*a-c*) and pheromone distribution (*b-d*) for $v=0$ (*a-b*) (the valley is constantly at $B=[0.0;0.0]$) and for $v=0.5$ (*c-d*) (the valley is traveling to southeast one cell for each two $t$ time step starting at the $B$ mid-point), for several time steps $t$ = 1,5,10,15,20,30,50,70 and 100.

When after some time steps the target is captured again the 2 clouds of agents congregate in one big cluster, and the chase continues with good tracking results. Somehow this emergent short path strategy is followed by the swarm, splitting in different clouds when needed for test speeds equal to $v=2$ or above that value (figs. 6,7). For dramatic values of speed however ($v=5$), the swarm is constantly and properly following the target but due to the value of severe speed, he is loosing the run. In order to overcame this lack of control the swarm itself self-regulates its population, and for dramatic speed tests of $v=10$ (fig.7c) or above, he explodes its population, trying to collect much spatial information as possible. Figure 8 can help us understand more profoundly the swarm behavior facing these different speeds. Fig. 8 (1st row) shows the population size as time passes.

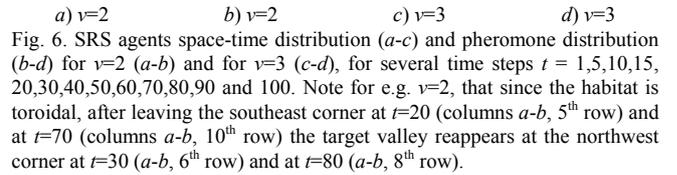

*a)* $v=2$     *b)* $v=2$     *c)* $v=3$     *d)* $v=3$

Fig. 6. SRS agents space-time distribution (*a-c*) and pheromone distribution (*b-d*) for $v=2$ (*a-b*) and for $v=3$ (*c-d*), for several time steps $t$ = 1,5,10,15, 20,30,40,50,60,70,80,90 and 100. Note for e.g. $v=2$, that since the habitat is toroidal, after leaving the southeast corner at $t=20$ (columns *a-b*, 5th row) and at $t=70$ (columns *a-b*, 10th row) the target valley reappears at the northwest corner at $t=30$ (*a-b*, 6th row) and at $t=80$ (*a-b*, 8th row).

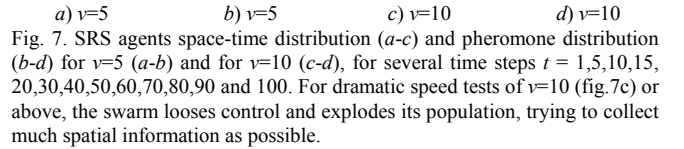

*a)* $v=5$     *b)* $v=5$     *c)* $v=10$     *d)* $v=10$

Fig. 7. SRS agents space-time distribution (*a-c*) and pheromone distribution (*b-d*) for $v=5$ (*a-b*) and for $v=10$ (*c-d*), for several time steps $t$ = 1,5,10,15, 20,30,40,50,60,70,80,90 and 100. For dramatic speed tests of $v=10$ (fig.7c) or above, the swarm looses control and explodes its population, trying to collect much spatial information as possible.

Fig. 8. Population size versus time (1st row) and mean altitude (averaged fitness) versus time (2nd row), for several runs under different environmental speeds: $v=0$, 0.5, 1, 1.5, 2, 3, 5 and $v=10$. Check the space-time distribution of these ant agents and their pheromone allocation (*social cognitive map* or societal memory [12,49,50]) in figs. 4-5-6-7.

Fig. 9. Mean Altitude (averaged fitness) versus time for several runs. For comparison purposes, we plotted bold curves for static *Ackley* functions with respective global minimum at *A*(-1.5;1.0), *B*(0.0;0.0) and *C*(1.0;-1.5). The other curves are for different landscape change updated frequencies: *uf*=50, 25, 10 and 5.

self-regulates under a common value, indicating a possible good tracking behavior accompanied by agent's specialization (exploitation nature). On the other hand, figure 8 (2nd row) shows what happens to mean agent's altitude over the *Ackley* landscape (*F* value on Eq. 8) as time go by. For speed tests up to *v*=2, the swarm can without difficulty track the extrema. For speed tests above that value, the behavior is more oscillatory, since at the toroidal borders the function assumes - due to simple and normal constraints in our representation - abrupt values from one border cell to their neighbors on "the other side". While the extrema is reasonably well tracked, ant-like agents very near the minimal extrema (near borders) could easily have high altitude "on the other side". Thus, even if the extrema is well tracked, mean altitude values achieve a visible increment. In addition, the swarm reaction speeds are undeniably very fast and the mean altitude quickly decreases often to values equal of those obtained with test speeds less or equal than *v*=2. This is particularly evident if we conduct a different and more dramatic test. In figure 9 we have updated the changes in the environment for different updated frequencies *uf*. For *uf*=5 (worst case) the base function minimal target point changes abruptly from *B* to *C*, from *C* to *A*, and from *A* to *B*, etc, at every 5 time steps. We see that even for this dramatic *uf* values, the swarm as a whole can still obtain performances, which in some cases outperform those obtained for static environments (bold curves). Reactions speeds are also quite high, generally in the range order of four to nine time steps.

The *SRS* performance can also be measured and analyzed taking in account the best minimum value found among all agents at each time step (zero is equivalent to full performance or the perfect capture of the right extrema). In figure 10 we have plotted these values for different test speeds. As expected, as speed increases the swarm passes more time away from the right extrema or close to it. The swarm however is able to capture the right extrema in perfection, for the most part of the run-time (the oscillatory behavior is again mainly due to our toroidal borders). From these results we can obtain the *SRS* success rate (fig. 11), defined as the number of times the colony as a whole was able not only to track, but to capture the right target over a run of 100 time steps, for each test speed. Remarkably the swarm is generally able to capture the right perfect extrema around 65% of the time, up to environmental speeds of 10 times the speed of a single ant-like agent or even for speeds greater than those values. Finally, very similar results were found for circular dynamics as introduced on section I.B. Random dynamics were not tested.

### B. Speed Tests and Severity

As discussed earlier in section I.B, *Angeline* [1] as well as *Bäck* [61,62], *Branke*, *Schmeck* [8] and others, studied not only how adaptive algorithms behave in drastic changing environments (as in the preceding subsection), but also on how the degree of these changes affect different approaches.

| | |
|---|---|
| $v = 0.0$ | $v = 2.0$ |
| $v = 0.5$ | $v = 3.0$ |
| $v = 1.0$ | $v = 5.0$ |
| $v = 1.5$ | $v = 10.0$ |

Fig. 10. Best minimum value found by *SRS* among all agents at each time step (zero is equivalent to full performance or the perfect capture of the right extrema). Plots are for different test speeds.

Fig.11. Averaged success rate (10 runs for each test speed). The success rate indicates the number of times the colony as a whole was able to track the right target over a run of 100 time steps, at each speed.

| | |
|---|---|
| $\delta$=0 (e.g. *s*=0.1,*T*=0 or *s*=1,*T*=0) | $\delta$=0.5 (e.g. *s*=0.1,*T*=5) |
| $\delta$=0.1 (e.g. *s*=0.1,*T*=1) | $\delta$=0.6 (e.g. *s*=0.1,*T*=6) |
| $\delta$=0.3 (e.g. *s*=0.1,*T*=3) | $\delta$=0.7 (e.g. *s*=0.1,*T*=7) |

Fig. 12. Sketches from the *Schaffer F7* complex multimodal function seen for different values of *T* related to the severity tests (here, *s*=0.1).

The question is important, since can help us understand on how strongly the system changes are and how severe this change is going to constrain the search nature of an algorithm. For these types of tests as well as for comparison purposes we have decided to implement a test function used recently by *Huang* and *Rocha* [56,30,29] with a Co-Evolutionary agent-based model of Genotype Editing (*ABMGE*).

This testbed is a dynamic version of the modified *Schaffer's F7* function studied in [29]. Several sketches of this multimodal function for different parametric values (different instances in time) are illustrated in fig. 12. Being *s* representing a parameter to set the severity and $X_i = x_i + \delta(t)$, with $-1 \leq x_i \leq 1$ for $i = 1,2$, a possible test problem [30] can be described by Equation 9:

$$f\left(\vec{X}\right) = 2.5 - \left(X_1^2 + X_2^2\right)^{0.25}\left[\sin^2\left(50\cdot\left(X_1^2 + X_2^2\right)^{0.1}\right) + 1\right] \quad (9)$$

In here, *X* and *Y* axis represent the index of the sample points in parameters $x_1$ and $x_2$ that are used to compute *f*(*x*), which is then represented on *z* axis, being our aim to maximize it. The example uses linear dynamics with severity *s*:

$$\delta(0) = 0,$$
$$\delta(T) = \delta(T-1) + s \quad (10)$$

Fig. 13. Best-so-far performance of a standard *GA* and the co-evolutionary *ABMGE* proposed by *Rocha* and *Huang* [56] on the linear dynamic *Schaffer F7* function with *s*=0.1 and *uf*=50, achieved by *Huang* and *Rocha* in [30].

Fig. 14. Best-so-far performance (best maximum found among all agents at each iteration) of the self-regulated swarm *SRS* approach against different severity values *s*=0.1, *s*=0.2, *s*=0.3, *s*=0.5, *s*=1.0, and *s*=1.5 (*uf*=50).

Fig. 15. *SRS* swarm population size at each iteration, for tests in figure 14.

As explained in [30], note that *T* is used as index for the environmental state; whenever the environment changes (e.g., every 50 generations in [30]), *T* is increased by 1. With these settings, *f(x)* has an optimal value of 2.5 among all ranges of *s*, here tested (this happens for *s*=0.1). This equation (10) will be used for the dynamic test function studied in the next section (IV), as well. In here we tested the *SRS* algorithm against severity values of *s*=0.1, *s*=0.2, *s*=0.3, *s*=0.5, *s*=1.0, and *s*=1.5 (for a constant updated frequency, *uf* = 50), as well as for different updated frequency's *uf* = 50, 25, 10 and 5 (worst dramatic case), for a constant severity of *s*=1. *SRS* parameter values were the usual ones as indicated in table 1, section II. Figure 13 displays the results achieved by *Huang* [30]; averaged best-so-far performance for the *Schaffer's F7* (Eq. 9) dynamic problem, with traditional genetic algorithms as well as with the co-evolutionary RNA editing model (*ABMGE*) [56], for a single test with *s*=0.1 and *uf*=50. As outlined by the authors, these results are encouraging since the co-evolutionary *ABMGE* model consistently outperforms the traditional GA in tracking the extrema. Meanwhile, figure 14 displays the results achieved by *SRS* against not only *s*=0.1 as well as for other values of severity *s* (all with *uf*=50). We see that the self-regulated swarm is not only able to track perfectly the correct extrema, nearly at all iterations for *s*=0.1, as achieves similar results for increasing values of severity. Please note that when *s* increases, equation 10 returns a different domain for our test dynamic function (different sample points for different *T*), thus the right respective extrema have optimal values less and obviously different than those present for *s*=0.1 (*z*=2.5). Meanwhile, figure 15 displays the *SRS* population size over the same tests. Since generally, the optimal peak form is the same (due to translation), for different severities (which did not happen in section III.A tests), we now see *SRS* self-regulating its population size similarly for different test runs (*SRS* tend to converge his population to values equal of those optimal and near optimal areas; circular concentric regions in fig. 12). As usual, *SRS* tends also to use more exploratory resources as severity increases.

Finally, we tested *SRS* against different updated frequencies. Respective results of performance and population size can be seen in figures 16 and 17. We see that even for the worst case scenario (*uf*=5 and *s*=1), *SRS* swarms are able to track the correct extrema in the majority of the test run-time. Naturally, there are oscillatory behaviors when the frequency updates, but again the swarm reaction speed is quite fast: in the order of ten to fifteen time-steps. As usual, at these kind of critical periods (environmental phase transitions) the swarm explodes his population (fig. 17), in order to overcome the abrupt changes in the landscape, using intelligently more resources to explore the novel sudden *habitat*. When again the extrema is correctly tracked after ten to fifteen time-steps, *SRS* self-regulates the number of their agents, downsizing it. Indeed, after the initial shock the resource economy is now so dominant and wise, that for certain cases the population decreases about nine times their earlier maximum size, settling down only at the correct peak.

Fig. 16. Best-so-far performance (best maximum found among all agents at each iteration) of the self-regulated swarm *SRS* approach against different updated frequencies *uf*=50, 25, 10 and 5 (*s*=1).

Fig. 17. *SRS* swarm population size at each iteration, for tests in figure 16.

## IV. DYNAMIC OPTIMAL CONTROL AND PROBLEM SOLVING BY EMERGENCE

Optimal Control Theory [34,70] aims to determine the control signals that will cause a process to satisfy the physical constraints and at the same time minimize (or maximize) some performance criterion. Practically, this kind of nonlinear and multiple local optimal problems (generally) appear in all engineering and science fields, and have been well studied from both theoretical and computational perspectives [30]. Once more, for comparison purposes we adopt an earlier artificial optimal control problem designed in [29], further studied and reported in [30]. The constraints of the artificial optimal control problem are:

$$\frac{d^2z(t)}{dt^2} + \sin(z(t))\frac{dz(t)}{dt} + \sin(t)\cos(z(t))z(t)^3 = \quad (11)$$
$$= \sin(t)u_1^2 + \cos(t)u_2^2 + \sin(t)u_1.u_2,$$
$$z(t_o) = 2, z'(t_0) = 2, t \in [0,1]$$

$\delta$=0 (e.g. *s*=1,*T*=0 or *s*=0.1,*T*=0) → region *A* in fig.12.

$\delta$=2 (e.g. *s*=1,*T*=2) → region *C* in fig.12.

$\delta$=3 (e.g. *s*=1,*T*=3) → region *D* in fig.12.

$\delta$=4 (e.g. *s*=1,*T*=4) → region *E* in fig.12.

$\delta$=6 (e.g. *s*=1,*T*=6) → region *G* in fig.12.

$\delta$=7 (e.g. *s*=1,*T*=7) → region *H* in fig.12.

Fig. 18. Sketch of the dynamic optimal control problem. Please note that the vertical $z(t_f)^2$ scale has changed for sketches *T*=6,7 since the highest peak for *T*=6 has now values slightly above 300, and slightly above 650 for *T*=7. An animated GIF can be found in [26].

where $U_i = u_i + \delta(t)$, with $-5 \leq u_i \leq 5$ for $i = 1,2$. As in [30], two examples are studied in this subsection that use linear dynamics with severity *s*=0.1 and *s*=1, according to equation 10 in the previous subsection, respectively. The goal is to maximize $z(t_f)^2$ by searching for two control variables, $u_1$ and $u_2$ (-5 ≤ $u_1$, $u_2$ ≤ 5). Sketches of this dynamic functions for $\delta$=0, 2, 3, 4, 6 and 7 are respectively illustrated in figure 18. *X* and *Y* represent the index of sample points in parameters $u_1$ and $u_2$, that are used to compute $z(t_f)^2$, which is then plotted on the Z-axis. In order to compute $z(t_f)^2$, the 2nd order differential equation (*DE*) in (11) must be converted into an equivalent system of two 1st order ordinary differential equations. The standard optimal control problem should now be represented by *max J($u_1$, $u_2$)* = $z(t_f)^2$, with $t_f$ = 1, $z(t_0)$=2, $y(t_0)$=2, $dz(t)/dt = y(t)$, and $dy(t)/dt$ given by Eq. 11. In order to numerically solve it for the different range values of $\delta$ (see Eq. 10), we have made use of a MATLAB version, working in reverse communication, of the Ordinary Differential Equation (*ODE*) solver DOPRI853 developed and coded in FORTRAN by *Hairer* et al. [24].

As in [30] we have completed tests for 400 iterations, with *s*=0.1 and *s*=1. Every 50 generations (*uf*=50), *T* (Eq. 10) is increased by 1.

*a*) *s*=0.1  *b*) *s*=0.1  *c*) *s*=1.0  *d*) *s*=1.0

Fig. 19. SRS agents space-time distribution (*a-c*) and pheromone distribution (*b-d*), with *uf*=50, for *s* = 0.1 (*a-b*) and for *s* = 1 (*c-d*). Maximal regions travel in waves from southeast to northwest. Figures are for several time steps *t* = 1, 25, 75, 100, 125, 150, 200, 250, 300, 350,375,400. An animated GIF can be found in [27].

The *SRS* self-regulated swarm space-time agent distribution and respective pheromone distribution are displayed in figure 19. We have used *SRS* values of Δ*e*=0.01 (remaining parameter values were the usual ones as indicated in table 1, section II). Figures 20 and 22, respectively show traditional *GA* and *ABMGE* results for *s*=0.1 and 1, achieved by *Huang* and *Rocha* [30,56]. They show that RNA editors [56], provide adaptive advantage for the ABMGE approach in tracking the extrema, while comparing it with traditional *GA*s.

Fig. 20. Best-so-far performance of a standard *GA* and the co-evolutionary *ABMGE* proposed by *Rocha* and *Huang* [56] on the dynamic optimal control function (11) with *s*=0.1 and *uf*=50, achieved by *Huang* and *Rocha* in [30].

Fig. 21. Best-so-far performance (best maximum found among all agents for each iteration) of the self-regulated swarm *SRS* approach on the dynamic optimal control function (11) with *s*=0.1 and *uf*=50.

On the other hand, figures 21 and 23, respectively show *SRS* results for *s*=0.1 and 1, achieved by our self-regulated swarm, clearly outperforming not only traditional *GA*s as the *ABMGE* model when facing dynamic environments. An exception however, should be made for *t* → [160,200] over the *s*=0.1 severity test, were *SRS* population exploded, due to sudden upcoming novel peak values over the domain. Figure 24 shows *SRS* population size as time passes, for the *s*=0.1 and *s*=1 tests. Generally, the standard *GA* and *ABMGE* does not have time, within 50 generations to exploit better peaks. As we can see from these pictures see the *SRS* reaction speed is generally higher, maintaining good results for different time-steps.

Fig. 22. Best-so-far performance of a standard *GA* and the co-evolutionary *ABMGE* proposed by *Rocha* and *Huang* [56] on the dynamic optimal control function (11) with *s*=1 and *uf*=50, achieved by *Huang* and *Rocha* in [30].

Fig. 23. Best-so-far performance (best maximum found among all agents for each iteration) of the self-regulated swarm *SRS* approach on the dynamic optimal control function (11) with *s*=1 and *uf*=50. Regions A,B, …, H correspond to regions in figure 18.

Fig. 24. Population size as time passes, of the self-regulated swarm *SRS* approach on the dynamic optimal control function (11) with *s*=0.1 and *s*=1 (*uf*=50).

## V. Discussion

### A. Emergence of a Societal Memory and Intelligent Path Planning via Stigmergy

As used in our *Self-Regulated Swarms* algorithm (*SRS*), stigmergy as a coordination mechanism is characterised by a lack of planning using implicit undirected communication between entities, a fact that makes it extremely flexible and robust for the exploration of large medium. The main idea relies on a form of asynchronous interaction and information interchange between entities mediated by an "active" environment. What characterizes stigmergy from other means of communication is, (1) the physical nature of the information released by the communicating insects, which corresponds to a modification of physical environmental states visited by the insects, and (2) the local nature of the released information, which can only be accessed by insects that visit the state in which it was released (or some neighbourhood of that state).

Agents engage on what is known as a *perception-action loop* [36]. Actions in the environment can influence agent's sensors, creating a loop. Relevant information flows from the environment via sensors to actuators, thus connecting them. The perception-action loop is important for understanding of adapted agents and to capture certain phenomena, like imprinting information onto the environment, offloading and later reacquisition of information. If we view sensors as capturing information and agent actuators as being capable of imprinting information on the dynamic environment, we can then treat agents as creating, maintaining and using various information flows, both internal and external. The view may be quite useful since there are strong indications that biological agents are partly driven by the necessity of acquiring, exchanging and also concealing information [36]. One of the possibilities is *uncovering hidden information*, which was demonstrated by *Kirsh* and *Maglio* [35] using the perception-action loop. In this work [35], advanced Tetris game players are shown to quickly rotate a falling block while it still is not visible completely. This active modification of the environment allows the players to discover the shape of the block before it actually becomes completely visible on the screen. Other possibility is *offloading of information* into the environment. This can be illustrated by the fact that we often write notes or reminders, which we then later look at to reacquire some information. A good account of such information flow between several people is given in the analysis of how members of an airliner crew indirectly communicate using cockpit controls as a medium [31]. For example, long before landing one of the pilots calculates proper flap settings for various speeds and then based on the results sets special markers on the airspeed indicator. Later, during the landing phase, the markers allow the crew to quickly and reliably find out what flap settings to use for the momentary speed. In a wider context, indirect communication via the environment is of high importance in distributed systems. Flexibility is possible, since environmental perturbations are directly connected to ant's behaviour: when the environment changes because of an external perturbation, the insects respond *appropriately* to that perturbation, as if it were a modification of the environment caused by the colonies activities. In other words, the colony can collectively respond to the perturbation with individuals exhibiting the same behaviour. When it comes to artificial agents, this type of flexibility is priceless: it means that agents can respond to a perturbation without being reprogrammed to deal with that particular instability. Over our dynamic tracking context, this means the possibility of achieving highly adaptive responses. Ants change the perceived environment of other ants (their cognitive map, according to *Chialvo* and *Millonas* [12,42,43]), and in every example, the environment serves as medium of communication. The problem in itself (dynamic environments) is implicitly used as a form of communication between agents trying to solve the problem. The environment is thus used as a global meta external memory, which is changing over time. This implicit use of the environment as memory allows ants to produce certain behaviour as a consequence of the effects produced in the local environment by previous behaviour. This meta global memory can be useful in cases where, for instance, a extrema reappears at a previous location or near from it, or when there is a necessity to maintain diversity. In the former case this environmental distributed implicit memory could recall that location and instantaneously move the population to the new optimum. As was pointed by *Branke* and *Schmeck* [8], it might also guide system's evolution to promising areas after a re-initialization (if adopted). But while the memory might allow exploitation of knowledge gained in the past, it might as well mislead evolution and prevent it from exploring new regions and discovering of new extrema [8]. Nevertheless, we believe that the positive feedback (allocation of pheromone) / negative feedback (evaporation) based stigmergic-reproductive highly adaptive approach implement on our proposal helps reducing that risk. For instance, an incoming new and bigger peak to the environment, even if initially yet not recognized, may contribute to higher altitude on agents near by that region peak, thus triggering a major evaporation rate on other regions, and pressuring those agents to leave that old region or to die from it. On the other hand, *SRS* does not use explicit representation of goals, and the dynamics of group behaviour are emergent and self-organizing, allowing for the appearance of strong phase transitions in behaviour (fig. 25). Distributed cognition and intelligent path planning is one of these interesting behaviours. A good illustration of this behaviour (among others available from different tests in sections III and IV) can be accessed from a previous test [17] (section II-B). In here, swarms with varying population size were submitted to a particular test function. Simulating changes in the environment consisted on changing the task from minimization ($t<250$) to maximization ($t>250$). We see that swarm performance was convincing and reinforced the idea that the system is highly adaptable and flexible. Moreover we see the appearance of an emergent intelligent path planning. After minimization is completed ($t<250$), for $t>250$ swarms move from local optima to local optima peaks, avoiding completely valleys on their path. First the swarm divides into two exploratory clusters of agents, one taking a northeast direction and then dying out due to inappropriate local conditions, while the other moves first to northwest, taking advantage of some local peaks and then moving to northeast, from an intermediate phase to the last final optimum peak.

Fig. 25. The curve shows the averaged agent's *y* spatial coordinate as time passes, for the test realized in figure 2 (section II B) [17]. After $t=250$, the swarm keeps moving north, emerging an intelligent path planning and showing a strong behavioral phase transition. Grey curves show upper and lower bounds for the entire population, at each time step.

### B. Self-Regulated Swarms (SRS) and SOS

As we have seen on the introductory sessions, many authors have tried to implement algorithmic solutions via the introduction of memory. Memory may be provided in two general ways: *implicitly* by using redundant or emergent representations (as in our *SRS* case), or *explicitly* by introducing an extra memory and formulating strategies to deposit and retrieve solutions later. In [9], *Branke* have compared a number of ways to organize an explicit memory (over a search population approach), and examined the usefulness of memory in different environments. As a result, he concluded that explicit memory is only useful when the environment repeatedly returns to a small set of previously experienced solutions. As they advance in [8], although the memory / search population approach was quite successful, it soon became obvious that a strategy based on (explicit) memorization would be too restricted to adapt successfully to a wide range of experiments.

As an alternative, *Branke* and *Schmeck* [8] developed more recently an approach with multiple populations acting as *Self-Organizing Scouts* (SOS), watching over the changing landscape. The novel idea is not only interesting as there are some similarities to our *Self-Regulated-Swarm* (*SRS*) interesting to point out, and from which future Self-Organizing approaches could ideally be designed. The basic idea of *SOS* is that once a peak has been found (i.e. the population converged to a high-performance region), the population should split: a small fraction, called the "child population" should "watch" over that peak, while the remainder of the population ("base population") should spread out and continue searching for new peaks. When a watched peak moves, the child population may follow it through space, and even request reinforcement. Since the population size is limited, unpromising peaks may be abandoned. To answer questions like, how to determine peaks that justify a population split-off, when should sub-populations ask for request, how many individuals should stay at each peak, or when should a peak be abandoned, *Branke* and *Schmeck* borrowed some ideas from the *Forking Genetic Algorithm* (*fGA*) as proposed by *Tsutui* et al. [65].

From these global *SOS* features, we can already detect some prominent similarities with *SRS*, at the level of our main objectives (reinforcement, division of labor, the number of individuals at each prominent peak, removal of individual with low fitness, etc). However, there are many differences on how these goals are achieved. Even if both algorithms rely on principles of Self-Organization, *Self-Regulated Swarms* (*SRS*) achieve these goals mainly by emergence, and several decisions, like reinforcement of a particular region, the number of individuals at each peak, division of labor (population split –off) and individual removal from the global system, are completely done via implicit mechanisms, through pheromone local allocation and global evaporation (distributed cognitive map [12,50,49]), as well as on the reproductive capabilities of each foraging agent.

## VI. CONCLUSIONS

In order to overcome difficult dynamic environment extrema tracking, we have proposed a *Self-Regulated Swarm* (*SRS*) algorithm which hybridizes the advantageous characteristics of Swarm Intelligence as the emergence of a societal environmental memory or cognitive map via collective pheromone laying in the landscape (balancing the exploration/exploitation nature of the search strategy), with a simple evolutionary mechanism that trough a direct reproduction procedure linked to local environment features is able to self-regulate the above exploratory swarm population, speeding it up globally.

The different experiments carry out on sections III and IV, demonstrate that *SRS* is able for quick adaptive response, outperforming results not only with standard Genetic Algorithms, as well as comparing it with very recently successful approaches based on Co-Evolution. From the different *SRS* features, we highlight some successful behaviors found:

(1) *SRS* is able to maintain a number of different solutions while adapting to new peaks appearing in the landscape.

(2) The experiments show that the *SRS* approach is able to spontaneously create and maintain (for certain instable changing conditions), different subpopulations on different peaks, emerging different exploratory corridors with intelligent path planning capabilities. In addition, these population split-offs only occurs when needed. For instance, in between static conditions temporal windows (when the upgrade frequency determines that the changes will only occur after some $T$ time steps), the population quickly converges to a single exploratory cluster. When however, the changes arrive, and for certain environmental conditions, the swarm may split again in two or more subpopulations.

(3) For dramatic conditions, the swarm is able to request more exploratory agents, reproducing more and scaling up its population. After this is not necessary and redundant (a highly-performing peak is found) however, the swarm can self-regulate down their number, scaling it down, thus economizing its search resources.

(4) The approach allows self-organized injection of larger diversity when needed, as we can perceive from grey curves in fig. 25.

(5) For certain environmental conditions, the swarm not only is able to smoothly track the dynamic extrema, as for many time-steps over some different temporal windows, is able to capture it in perfection.

(6) Last but not least, we prove that our *SRS* collective swarm of ant-like agents is able to track about 65% of moving peaks traveling up to ten times faster than the velocity of a single ant composing that precise swarm tracking system. This emerged behavior is probably one of the most interesting ones achieved by the present work.

We expect the framework proposed in this work to advance the current state of research on Swarm Intelligence and Evolutionary Computation in dynamic Environments.

## ACKNOWLEDGMENT

The authors would like to thank to Professor *Fernando Durão* (CVRM-IST, Technical Univ. of Lisbon, Portugal) for providing the MATLAB version, working in reverse communication, of the Ordinary Differential Equation (ODE) solver DOPRI853 developed and coded in FORTRAN by *E. Hairer*, *S.P.Norset* and *G.Wanner* [24]. The second author wishes to thank FCT, *Ministério da Ciência e Tecnologia - Portugal,* for his research fellowship SFRH/BD/18868/2004.

## ORIGINAL PAPER WITH FULL PICTURES AVAILABLE IN:

http://alfa.ist.utl.pt/~cvrm/staff/vramos/Vramos-NIAS06.pdf